\documentclass[%
aps,
prl,
reprint,
superscriptaddress,
showpacs,preprintnumbers,
nobibnotes,
amsfonts,
amsmath,
amssymb,
]{revtex4-1}

\usepackage{graphicx}
\usepackage{dcolumn}
\usepackage{bm}
\usepackage{gensymb}

\usepackage{lineno}

\begin{document}


\title{Measurement of the Vector and Tensor Asymmetries at Large Missing \\
Momentum in Quasielastic $\bm{(\vec{e}, e^{\prime}p)}$ Electron Scattering from Deuterium}

\author{A.~DeGrush} \thanks{Results based on Ph.D. theses of
  A.~DeGrush and A.~Maschinot.} \affiliation{Laboratory for Nuclear Science and Bates Linear
  Accelerator Center, Massachusetts Institute of Technology, Cambridge, MA 02139} 
\author{A.~Maschinot}
\affiliation{Laboratory for Nuclear Science and Bates Linear Accelerator
  Center, Massachusetts Institute of Technology, Cambridge, MA 02139} 
\author{T.~Akdogan}
\affiliation{Laboratory for Nuclear Science and Bates Linear Accelerator
  Center, Massachusetts Institute of Technology, Cambridge, MA 02139} 
\author{R.~Alarcon}
\affiliation{Arizona State University, Tempe, AZ 85287}
\author{W.~Bertozzi}
\affiliation{Laboratory for Nuclear Science and Bates Linear Accelerator
  Center, Massachusetts Institute of Technology, Cambridge, MA 02139} 
\author{E.~Booth}
\affiliation{Boston University, Boston, MA 02215}
\author{T.~Botto}
\affiliation{Laboratory for Nuclear Science and Bates Linear Accelerator
  Center, Massachusetts Institute of Technology, Cambridge, MA 02139} 
\author{J.R.~Calarco}
\affiliation{University of New Hampshire, Durham, NH 03824}
\author{B.~Clasie}
\affiliation{Laboratory for Nuclear Science and Bates Linear Accelerator
  Center, Massachusetts Institute of Technology, Cambridge, MA 02139} 
\author{C.~Crawford}
\affiliation{University of Kentucky, Lexington, KY 40504}
\author{K.~Dow}
\affiliation{Laboratory for Nuclear Science and Bates Linear Accelerator
  Center, Massachusetts Institute of Technology, Cambridge, MA 02139} 
\author{M.~Farkhondeh}
\affiliation{Laboratory for Nuclear Science and Bates Linear Accelerator
  Center, Massachusetts Institute of Technology, Cambridge, MA 02139} 
\author{R.~Fatemi}
\affiliation{University of Kentucky, Lexington, KY 40504}
\author{O.~Filoti}
\affiliation{University of New Hampshire, Durham, NH 03824}
\author{W.~Franklin}
\affiliation{Laboratory for Nuclear Science and Bates Linear Accelerator
  Center, Massachusetts Institute of Technology, Cambridge, MA 02139} 
\author{H.~Gao}
\affiliation{Triangle Universities Nuclear Laboratory and Duke University,
  Durham, NC 27708}
\author{E.~Geis}
\affiliation{Arizona State University, Tempe, AZ 85287}
\author{S.~Gilad}
\affiliation{Laboratory for Nuclear Science and Bates Linear Accelerator
  Center, Massachusetts Institute of Technology, Cambridge, MA 02139} 
\author{D.K.~Hasell}
\email[Correspondence: ]{hasell@mit.edu}
\affiliation{Laboratory for Nuclear Science and Bates Linear Accelerator
  Center, Massachusetts Institute of Technology, Cambridge, MA 02139}
\author{P.~Karpius}
\affiliation{University of New Hampshire, Durham, NH 03824}
\author{M.~Kohl}
\affiliation{Hampton University, Hampton, VA 23668 and Thomas Jefferson
National Accelerator Facility, Newport News, VA 23606}
\author{H.~Kolster}
\affiliation{Laboratory for Nuclear Science and Bates Linear Accelerator
  Center, Massachusetts Institute of Technology, Cambridge, MA 02139} 
\author{T.~Lee}
\affiliation{University of New Hampshire, Durham, NH 03824}
\author{J.~Matthews}
\affiliation{Laboratory for Nuclear Science and Bates Linear Accelerator
  Center, Massachusetts Institute of Technology, Cambridge, MA 02139} 
\author{K.~McIlhany}
\affiliation{United States Naval Academy, Annapolis, MD 21402}
\author{N.~Meitanis}
\affiliation{Laboratory for Nuclear Science and Bates Linear Accelerator
  Center, Massachusetts Institute of Technology, Cambridge, MA 02139} 
\author{R.~Milner}
\affiliation{Laboratory for Nuclear Science and Bates Linear Accelerator
  Center, Massachusetts Institute of Technology, Cambridge, MA 02139} 
\author{J.~Rapaport}
\affiliation{Ohio University, Athens, OH 45701}
\author{R.~Redwine}
\affiliation{Laboratory for Nuclear Science and Bates Linear Accelerator
  Center, Massachusetts Institute of Technology, Cambridge, MA 02139} 
\author{J.~Seely}
\affiliation{Laboratory for Nuclear Science and Bates Linear Accelerator
  Center, Massachusetts Institute of Technology, Cambridge, MA 02139} 
\author{A.~Shinozaki}
\affiliation{Laboratory for Nuclear Science and Bates Linear Accelerator
  Center, Massachusetts Institute of Technology, Cambridge, MA 02139} 
\author{A.~Sindile}
\affiliation{University of New Hampshire, Durham, NH 03824}
\author{S.~\v{S}irca}
\affiliation{Faculty of Mathematics and Physics, University of Ljubljana,
  and Jo\v{z}ef Stefan Institute, 1000 Ljubljana, Slovenia}
\author{E.~Six}
\affiliation{Arizona State University, Tempe, AZ 85287}
\author{T.~Smith}
\affiliation{Dartmouth College, Hanover, NH 03755}
\author{B.~Tonguc}
\affiliation{Arizona State University, Tempe, AZ 85287}
\author{C.~Tschal\" ar}
\affiliation{Laboratory for Nuclear Science and Bates Linear Accelerator
  Center, Massachusetts Institute of Technology, Cambridge, MA 02139} 
\author{E.~Tsentalovich} 
\affiliation{Laboratory for Nuclear Science and Bates Linear Accelerator
  Center, Massachusetts Institute of Technology, Cambridge, MA 02139} 
\author{W.~Turchinetz} 
\thanks{Deceased}
\affiliation{Laboratory for Nuclear Science and Bates Linear Accelerator
  Center, Massachusetts Institute of Technology, Cambridge, MA 02139} 
\author{Y.~Xiao}
\affiliation{Laboratory for Nuclear Science and Bates Linear Accelerator
  Center, Massachusetts Institute of Technology, Cambridge, MA 02139} 
\author{W.~Xu} 
\affiliation{Triangle Universities Nuclear Laboratory and Duke University,
  Durham, NC 27708}
\author{Z.-L.~Zhou}
\affiliation{Laboratory for Nuclear Science and Bates Linear Accelerator
  Center, Massachusetts Institute of Technology, Cambridge, MA 02139} 
\author{V.~Ziskin}
\affiliation{Laboratory for Nuclear Science and Bates Linear Accelerator
  Center, Massachusetts Institute of Technology, Cambridge, MA 02139} 
\author{T.~Zwart}
\affiliation{Laboratory for Nuclear Science and Bates Linear Accelerator
  Center, Massachusetts Institute of Technology, Cambridge, MA 02139}

\collaboration{The BLAST collaboration}

\date{\today}

\begin{abstract}
  We report the measurement of the beam-vector and tensor asymmetries
  $A^V_{ed}$ and $A^T_d$ in quasielastic $(\vec{e}, e^{\prime}p)$
  electrodisintegration of the deuteron at the MIT-Bates Linear
  Accelerator Center up to missing momentum of 500~MeV/c. Data were
  collected simultaneously over a momentum transfer range
  $0.1< Q^2<0.5$~(GeV/c)$^2$ with the Bates Large Acceptance
  Spectrometer Toroid using an internal deuterium gas target,
  polarized sequentially in both vector and tensor states.  The data
  are compared with calculations. The beam-vector asymmetry $A^V_{ed}$
  is found to be directly sensitive to the $D$-wave component of the
  deuteron and have a zero-crossing at a missing momentum of about
  320~MeV/c, as predicted.  The tensor asymmetry $A^T_d$ at large
  missing momentum is found to be dominated by the influence of the
  tensor force in the neutron-proton final-state interaction.  The new
  data provide a strong constraint on theoretical models.
\end{abstract}



\maketitle

Understanding the structure and properties of the nucleon-nucleon
system is a cornerstone of nuclear physics. Classic studies of the
properties of the bound state, (the deuteron) like the magnetic and
quadrupole moments, have elucidated the non-relativistic $S$- and
$D$-state wave function components.  However, modern polarized beams
and targets provide new tools to revisit this subject to provide more
stringent tests of our understanding.  Spin-dependent quasielastic
$(\vec{e}, e^{\prime}p)$ electron scattering from both vector and
tensor polarized deuterium provides unique access to the orbital
angular momentum structure of the deuteron, which is inaccessible in
unpolarized scattering~\cite{Boeglin:2014gda}. The combination of a
pure, highly polarized gas target internal to a storage ring with an
intense, highly polarized electron beam and a large acceptance
detector allows the simultaneous measurement of the asymmetries as a
function of initial-state proton momentum and momentum transfer. To
see the direct effects of the $D$-state, initial-state momenta up to
500~MeV/c are required.  Further, nucleon-nucleon correlations with
high relative momenta are known to play a significant role in nuclear
structure~\cite{Hen2016}.  The tensor force between the neutron and
proton can be probed via final-state interaction (FSI) effects in
spin-dependent quasielastic $^2{\rm H}(\vec{e},e^{\prime}p)$ at large
initial-state momenta~\cite{Jeschonnek:2017, Mayer:2016zmv}.  In this
Letter, we report on new measurements of the vector and tensor
asymmetries in quasielastic $(\vec{e}, e^{\prime}p)$ scattering from
deuterium over a broad range of kinematics and compare with
theoretical calculations.

The deuteron's simple structure enables reliable calculations to be
performed in sophisticated theoretical frameworks. These calculations
use nucleon-nucleon potentials as input, which show that the ground-state
wave function is dominated by the $S$-state at low relative
proton-neutron momentum ${\bm p}$.  The tensor component of the NN
interaction generates an additional $D$-state component.  Models
predict that the $S-$ and $D-$state components strongly depend on
${\bm p}$.  In the $^2{\rm H}(\vec{e}, e^{\prime}p)$ reaction, energy
$\nu$ and three-momentum $\bm q$ are transferred to the deuteron.  The
cross section can be measured as a function of the missing momentum
${\bm p_m} \equiv {\bm q} - {\bm p_f}$, where ${\bm p_f}$ is the
measured momentum of the ejected proton.

The cross section can be written in terms of the unpolarized
cross section $S_0$ multiplied by asymmetries diluted by various
combinations of the beam's longitudinal polarization $h$, the target
vector polarization $P_z$, and the target tensor polarization
$P_{zz}$~\cite{Aren1992} as:
\begin{eqnarray}
\frac{d\sigma}{d\omega d\Omega_e d\Omega^{CM}_{pn}}
&=&S_0\,[\,1+P_z A^V_d  + P_{zz}A^T_d \notag\\
&&\,+\,h\,(\,A_e+P_zA^V_{ed}+P_{zz}A^T_{ed}\,)\,]
\end{eqnarray}
In the Born approximation $A_e$, $A^V_d$, and $A^T_{ed}$ are all zero.
In a purely $S$-state $A^T_d$ is also zero but will vary from zero as
$D$-state contributions become important providing a measure of the
tensor component of the NN interaction. Similarly, $A^V_{ed}$ will
vary from $hP_z$ as $D$-state contributions become significant.

Previous measurements of the asymmetries $A^T_{d}$ up to $p_m=$
150~MeV/c~\cite{NIKHEF:1999} and of $A^V_{ed}$ up to $p_m=$
350~MeV/c~\cite{NIKHEF:2002} were carried out at NIKHEF.  These
pioneering measurements did not have the kinematic reach to observe
the effects of the $D$-state in $A^V_{ed}$ or the FSI effects in
$A^T_d$.

Our experiment was carried out with the Bates Large Acceptance
Spectrometer Toroid (BLAST)~\cite{Hasell:2009,Hasell:2011}.  The BLAST
experiment; including details on the detector, the South Hall Ring
(SHR) of the MIT-Bates Linear Accelerator Center, the longitudinal
polarized electron beam, the atomic beam source~\cite{Cheever:2006}
(ABS) that produced the vector and tensor polarized deuterium, and the
experimental operation; has been described extensively in the cited
references and will not be repeated here.

The target spin states were switched every 5~minutes. The longitudinal
beam polarization was reversed every injection cycle and was monitored
continuously using a Compton back-scattering polarimeter.  The average
polarization was $h=0.6558 \pm 0.007 {\rm (stat)} \pm 0.04 {\rm (sys)}$.

The data were taken in two separate running periods and acquired
simultaneously with the BLAST measurements of $G^n_E$~\cite{Geis:2008}
and $T_{20}$~\cite{Zhang:2011}. The average target spin angles were
$31.3\degree\pm0.43\degree$ and $47.4\degree\pm0.45\degree$ with
respect to the beam axis for the two run periods. The target spin
angle was in the horizontal plane pointing into the left sector and
was determined using elastic electron-deuteron
scattering~\cite{Zhang:2011}. Electrons scattered into the right
(left) sector delivered momentum transfer predominantly parallel
(perpendicular) to the target spin vector, so-called {\it same sector}
({\it opposing sector}) kinematics.

The average product of beam and target polarization was determined
from measuring $A^V_{ed}$ in the $^2\vec{\rm H}(\vec{e}, e^\prime p)$
reaction in the quasielastic limit (low missing momentum,
$p_m<0.1$~GeV/c) where the reaction is close to elastic $ep$
scattering.  The results were
$hP_z = 0.5796 \pm 0.0034 {\rm (stat)} \pm 0.0054 {\rm (syst)}$ in the
first run and $0.5149 \pm 0.0043 {\rm (stat)} \pm 0.0054 {\rm (syst)}$
in the second run. In parallel, $hP_z$ was similarly determined from
the quasielastic $^2\vec{\rm H}(\vec{e}, e^\prime n)$ reaction and was
found to be in good agreement.

The target tensor polarizations were determined from fits to the
elastic electron-deuteron observable $T_{20}$~\cite{Zhang:2011} using
parameterizations to previous data~\cite{Abbott:2000}.  The results
were $P_{zz} = 0.683 \pm 0.015 \pm 0.013 \pm 0.034$ and
$0.563 \pm 0.013 \pm 0.023 \pm 0.028$, where the three uncertainties
are statistical, systematic, and due to the parametrization of
$T_{20}$, in that order.

\begin{figure}[!t]
  \includegraphics[width=0.48\textwidth, viewport=7 8 1617
  1056,clip]{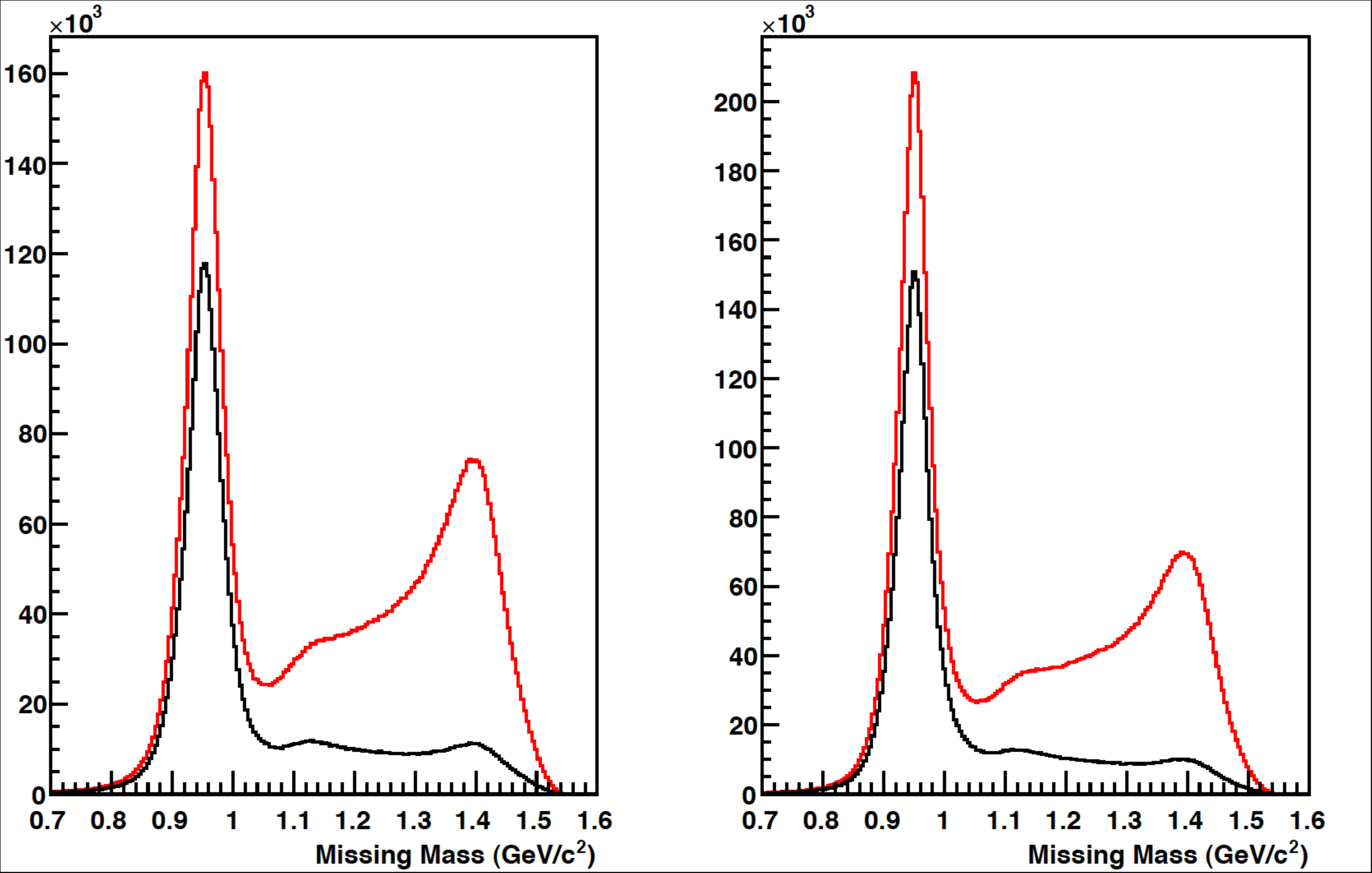}
  \caption{Histograms (color online) of the yields versus missing mass for target spin
    angle $\approx31\degree$ without (red) and with (black)
    \v{C}erenkov cuts for $0.1<Q^2<0.5$~(GeV/c)$^2$ for {\it opposing}
    (left) and {\it same} (right) sector kinematics.}
\label{fig:missingm}
\end{figure}

The event selection is described in detail in the theses of
A.~Maschinot~\cite{Maschinot} and A.~DeGrush~\cite{DeGrush}.  Briefly,
electron-proton coincidence events were selected using a series of
PID, timing, and vertex cuts. Events were chosen with two oppositely
charged (curvature) tracks in opposing sectors.  The \v{C}erenkov
detectors were used to distinguish electrons from $\pi^-$ and time of
flight was used to select proton events while rejecting events with
$\pi^+$ or a deuteron. To ensure that the two particles came from the
same event, a cut was placed on the relative separation of their
vertices in the target $|z_p - z_e| < 5$~cm. Once these events were
selected each track's kinematic variables,
$(p_e, \theta_e, \phi_e, z_e)$ for the electron and
$(p_p, \theta_p, \phi_p, z_p)$ for the proton, were used to determine
the variables $(Q^2, p_m, m_m)$. The quasielastic events were selected
by placing a $2.5\sigma$ cut around the peak of the missing mass
spectrum (see Fig.~\ref{fig:missingm}) representing the remaining
neutron.

\begin{figure*}[!ht]
\includegraphics[width=0.246\textwidth,viewport=0 45 562 690, clip]{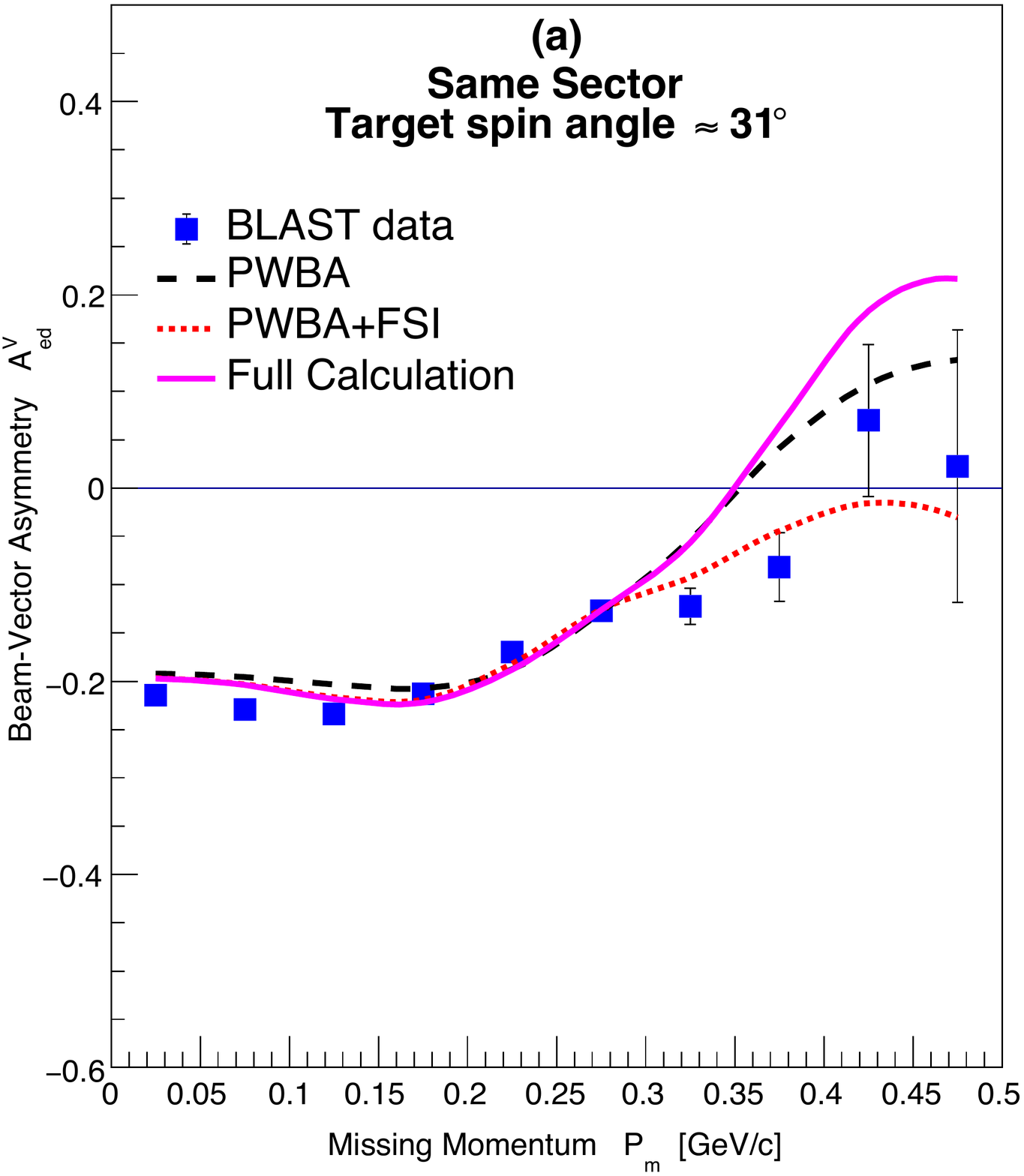}
\hfil
\includegraphics[width=0.246\textwidth,viewport=0 45 562 690, clip]{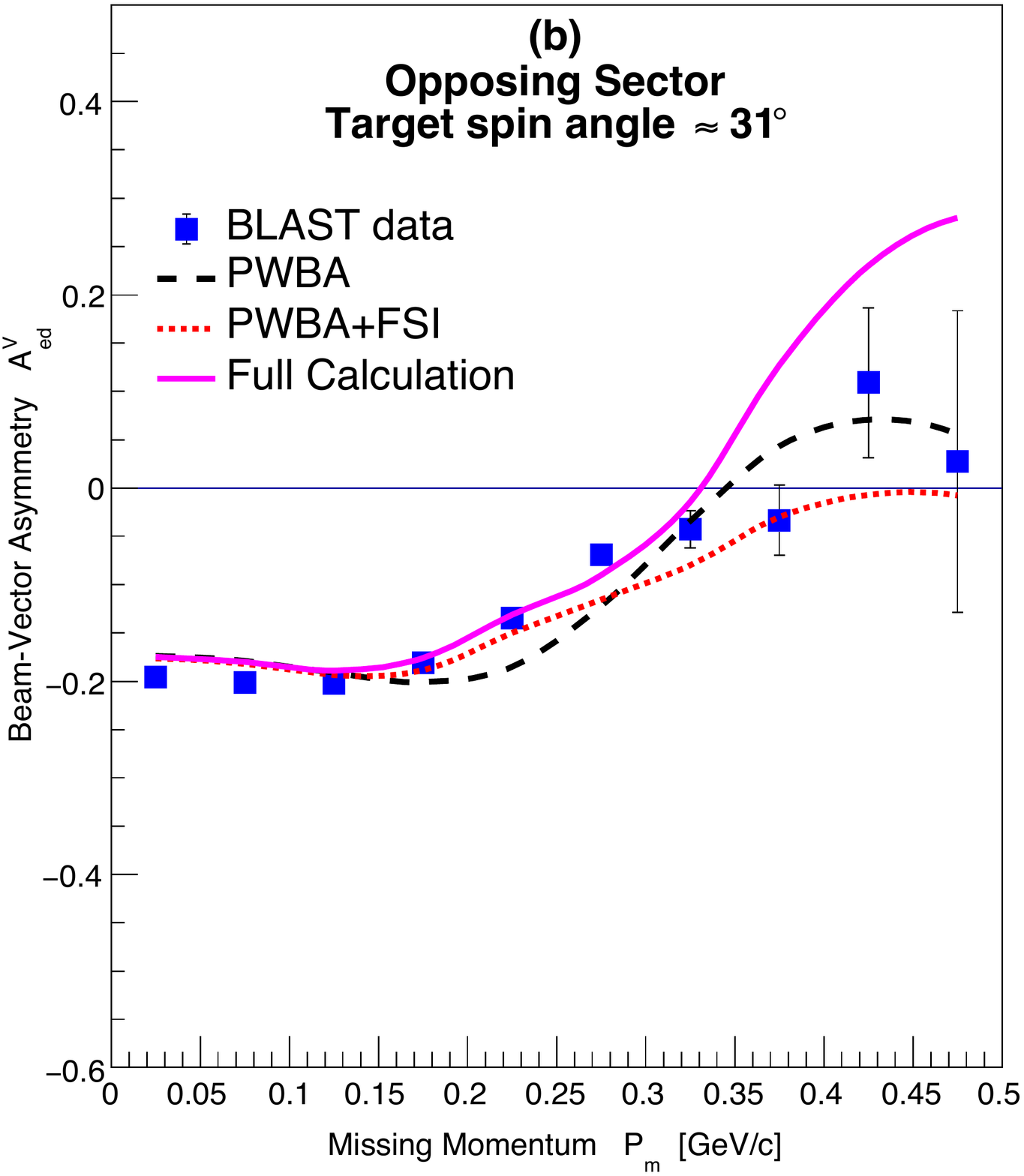}
\hfil
\includegraphics[width=0.246\textwidth,viewport=0 45 562 690, clip]{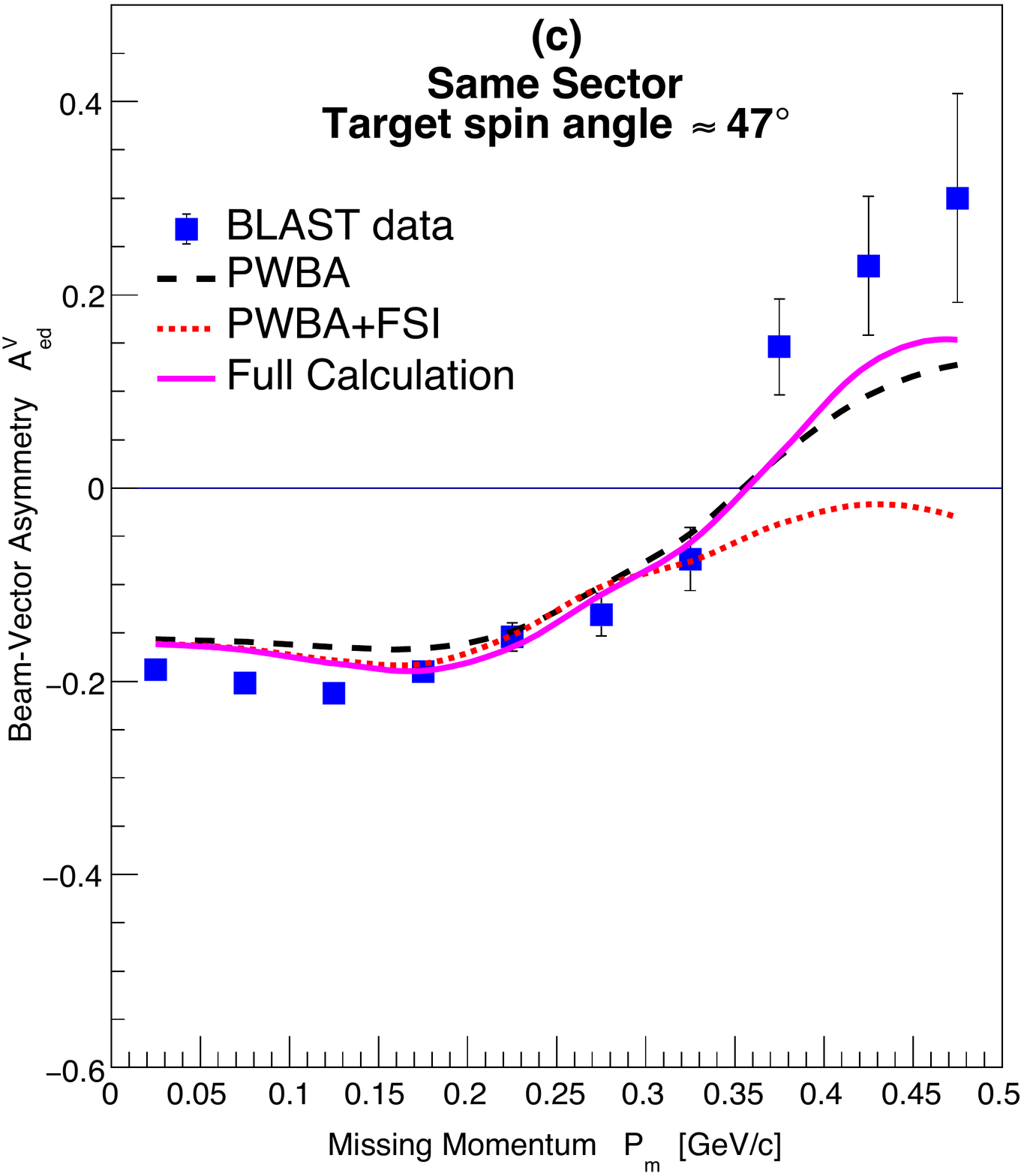}
\hfil
\includegraphics[width=0.246\textwidth,viewport=0 45 562 690, clip]{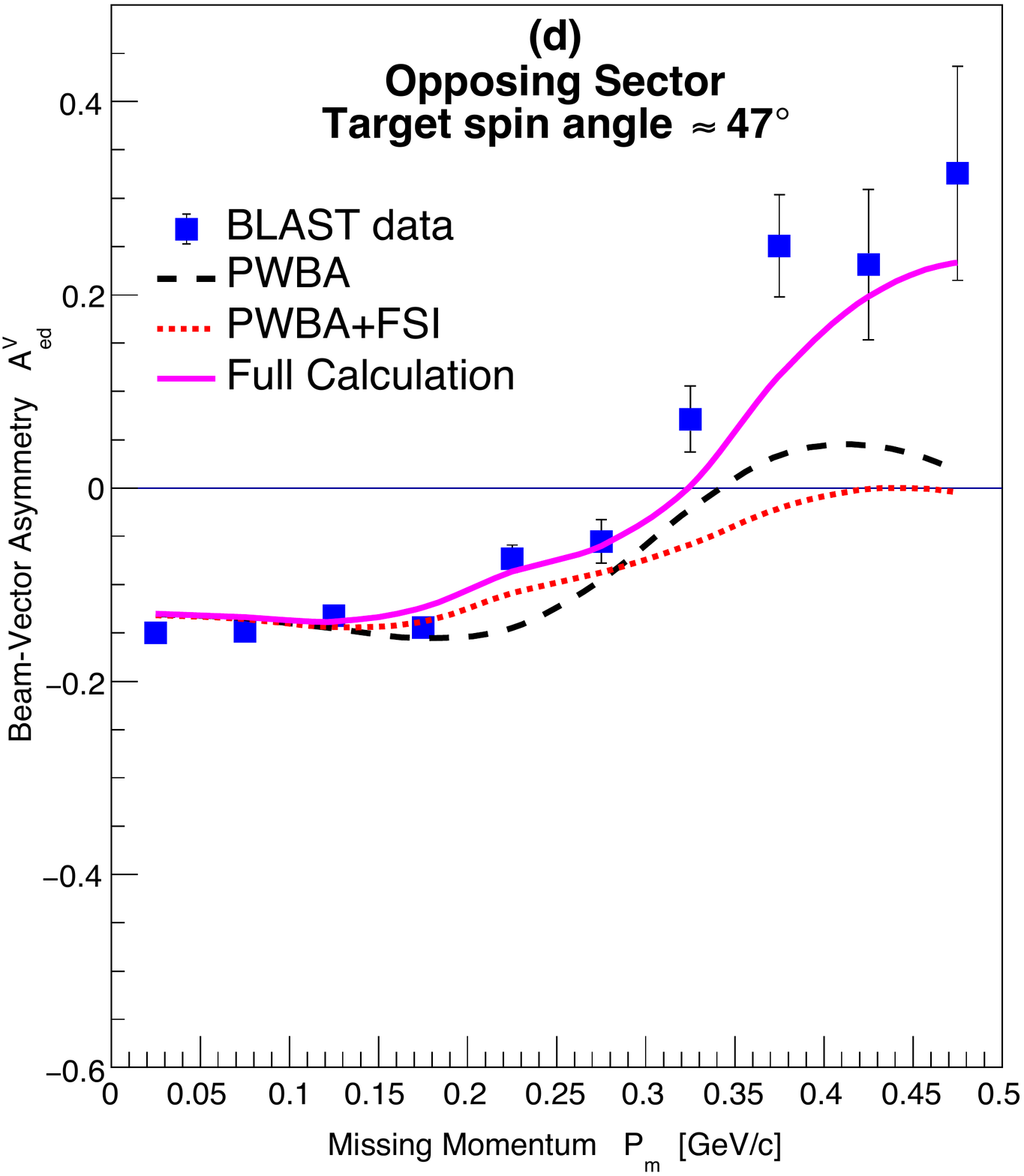}
\caption{Beam vector asymmetries $A^V_{ed}$ for
  $0.1< Q^2< 0.5$~(GeV/c)$^2$ vs. $p_m$.  Panels (a)
  and (c) refer to {\it same sector} kinematics for target spin angles
  $\approx31^\circ$ and $\approx47^\circ$.  Panels (b) and (d) refer
  to {\it opposing sector} kinematics for the same target spin
  angles.}
\label{fig:vector}
\end{figure*}
 
After background subtraction and correcting for false asymmetries
determined from the empty target runs the resulting yield in the
various $Q^2$ and $p_m$ bins could be determined for the combinations
of beam and deuterium vector and tensor orientations ($\pm$1, $\pm$1
or 0, +1 or -2) for which data was collected. The charge normalized
yields or event rates could be combined to give the desired
asymmetries.  For this paper:
\begin{eqnarray}
S_0&=&\frac{1}{6}[R(1, 1, 1)+R(-1, 1, 1)\notag\\
&&+R(1, -1, 1)+R(-1, -1, 1)\notag\\
&&+2R(1, 0, -2)+2R(-1, 0, -2)]\\
A^V_{ed}&=&\frac{1}{4hP_zS_0}[R(1, 1, 1)-R(-1, 1, 1)\notag\\
&&-R(1, -1, 1)+R(-1, -1, 1)]\\
A^T_d&=&\frac{1}{12P_{zz}S_0}[R(1, 1, 1)+R(-1, 1, 1)\notag\\
&&+R(1, -1, 1)+R(-1, -1, 1)\notag\\
&&-2R(1, 0, -2)+-2R(-1, 0, -2)]
\end{eqnarray}
where $R(h, P_z, P_{zz})$ is the charge normalized yield or event rate
for each spin orientation combination.

Radiative corrections to the asymmetries were calculated using the
MASCARAD code~\cite{MASCARAD} and all found to be less than 1\%. Thus,
no corrections were applied to the asymmetries but a systematic
uncertainty of $\pm1$\% was included. Background arose predominantly
from beam collisions with the target cell wall. Estimates for this
rate were made by acquiring data with and without gas in the target
cell. Background was subtracted on a bin-by-bin basis and increased
from a typical value of $<1$\% at low $p_m$ to of order 10\% at the
highest $p_m$.

The beam-vector asymmetries $A^V_{ed}$ for the runs with the two
target spin orientations are shown in Fig.~\ref{fig:vector}. The data
are shown in {\it same sector} and {\it opposing sector} kinematics as
a function of the missing momentum $p_m$ for momentum transfers
$0.1<Q^2<0.5$~(GeV/c)$^2$. The values of $p_m$ extend up to about
500~MeV/c and the data are compared with theoretical calculations
based on the model of Arenh\"ovel, Leidemann, and
Tomusiak~\cite{Aren2005}. The model was calculated for the kinematics
of the experiment folding in the detector acceptances and efficiencies
in a comprehensive GEANT simulation. The curves shown in each plot
correspond to a plane-wave Born approximation (PWBA) which includes
the coupling to the neutron, a PWBA with final state interactions
(FSI) and a full calculation beyond PWBA+FSI including the effects of
meson-exchange currents (MEC), isobar configurations (IC) and
relativistic corrections (RC). The two-body wave functions needed for
the calculation of the observables are based on the realistic Bonn
potential~\cite{Bonn1987}, which is defined in purely nucleonic space.
The theoretical calculations were found to be insensitive to the
choice of different realistic potentials ({\it e.g.}
Reid~\cite{Reid:1968}, Paris~\cite{Paris:1980}, Argonne V14 and
V18~\cite{Argonne:1995}). The treatment of MEC, IC, and RC is done
consistently according to~\cite{Aren2005,Bonn1997}.

At the $p_m = 0$ limit the opposing sector asymmetries are directly
proportional to the product $hP_z$, a key parameter that has been
determined with better than 1\% absolute accuracy. The target vector
polarization $P_z$ is directly related to the polarization $P$ of the
proton or neutron bound in the deuteron such that~\cite{Aren1988}
\begin{eqnarray}
P &=&\sqrt{\frac{2}{3}} P_z(P_S - \frac{1}{2}P_D)
\end{eqnarray} 
where $P_S$ and $P_D$ are the $S$- and $D$-state probabilities of the
deuteron, respectively.  This illustrates the fact that the
polarization of a nucleon in the $D$-state is opposite to that of a
nucleon in the $S$-state, as expected from angular momentum
considerations for a $J^{\pi} = 1^+$ system like the deuteron. The
present results for the $A^V_{ed}$ asymmetries show for the first time
the evolution going from the $S$-state to the $D$-state in momentum
space. The $A^V_{ed}$ are constant up to about $p_m =150$~MeV/c which
is consistent with an $S$-state, then as $p_m$ increases, the presence
of the $D$-state lowers the proton polarization in the deuteron until
it changes sign when $P_D\geq2P_S$.

\begin{figure*}[htbp]
\includegraphics[width=0.246\textwidth, viewport=0 45 562 690, clip]{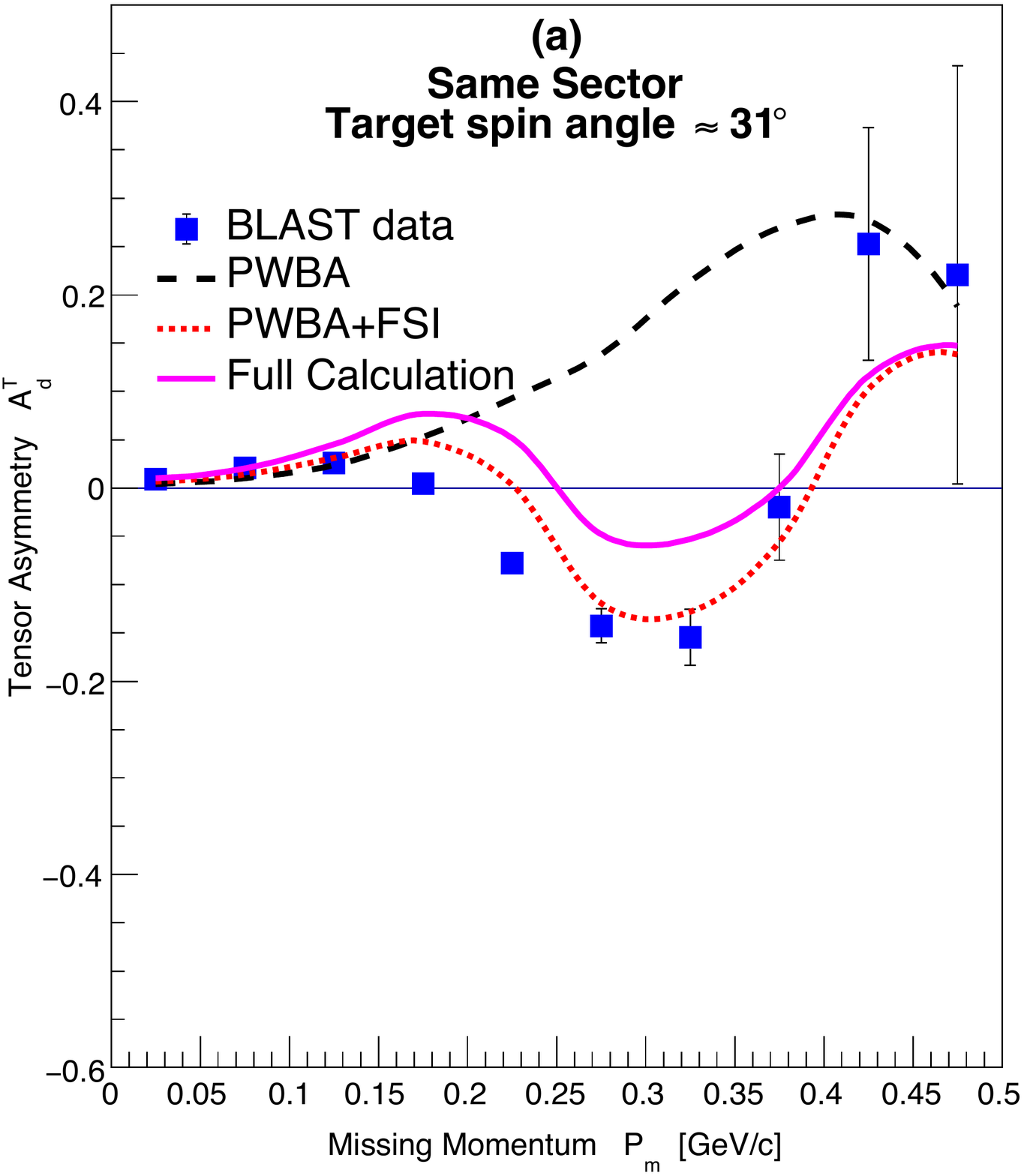}
\hfil
\includegraphics[width=0.246\textwidth, viewport=0 45 562 690, clip]{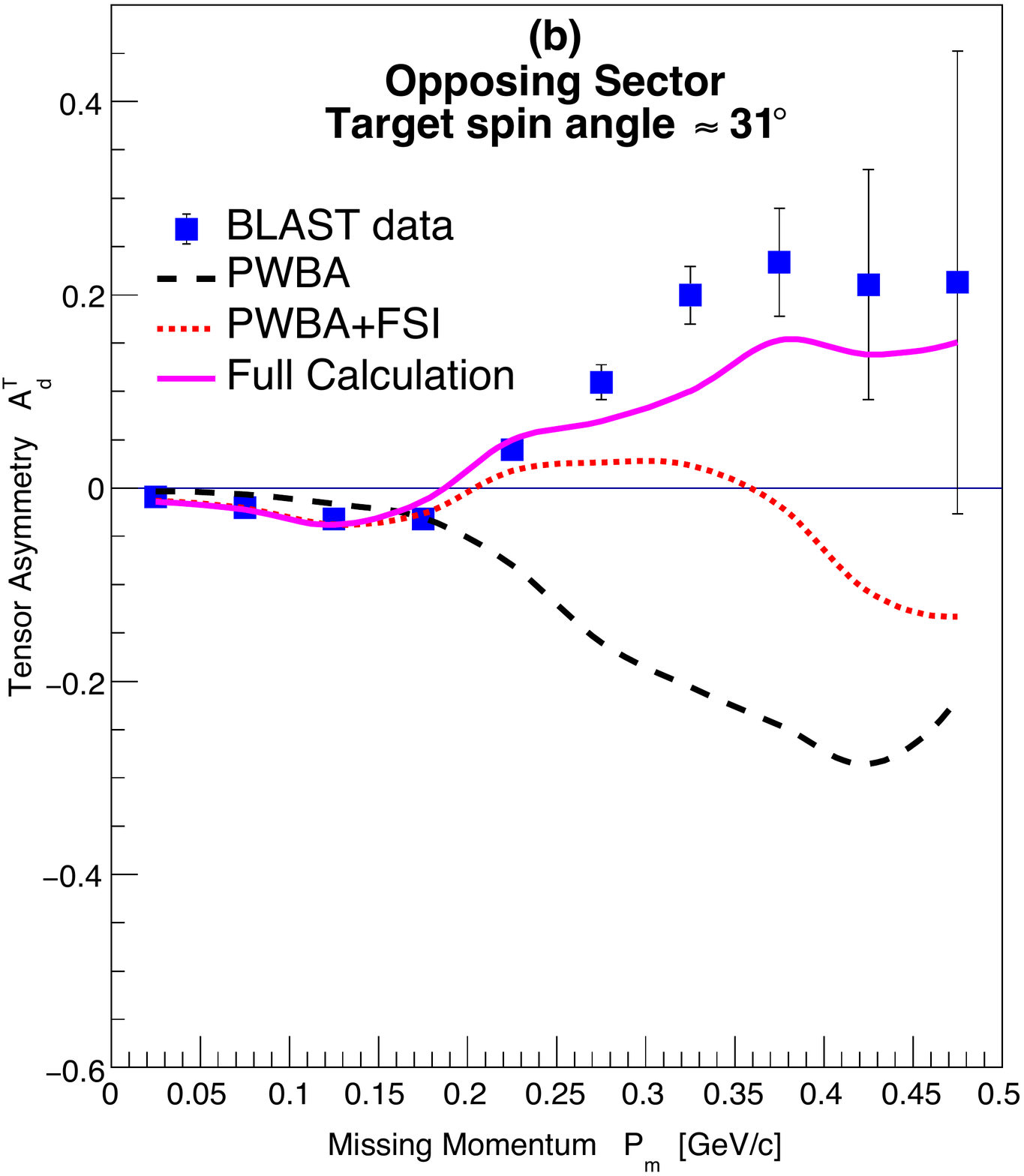}
\hfil
\includegraphics[width=0.246\textwidth, viewport=0 45 562 690, clip]{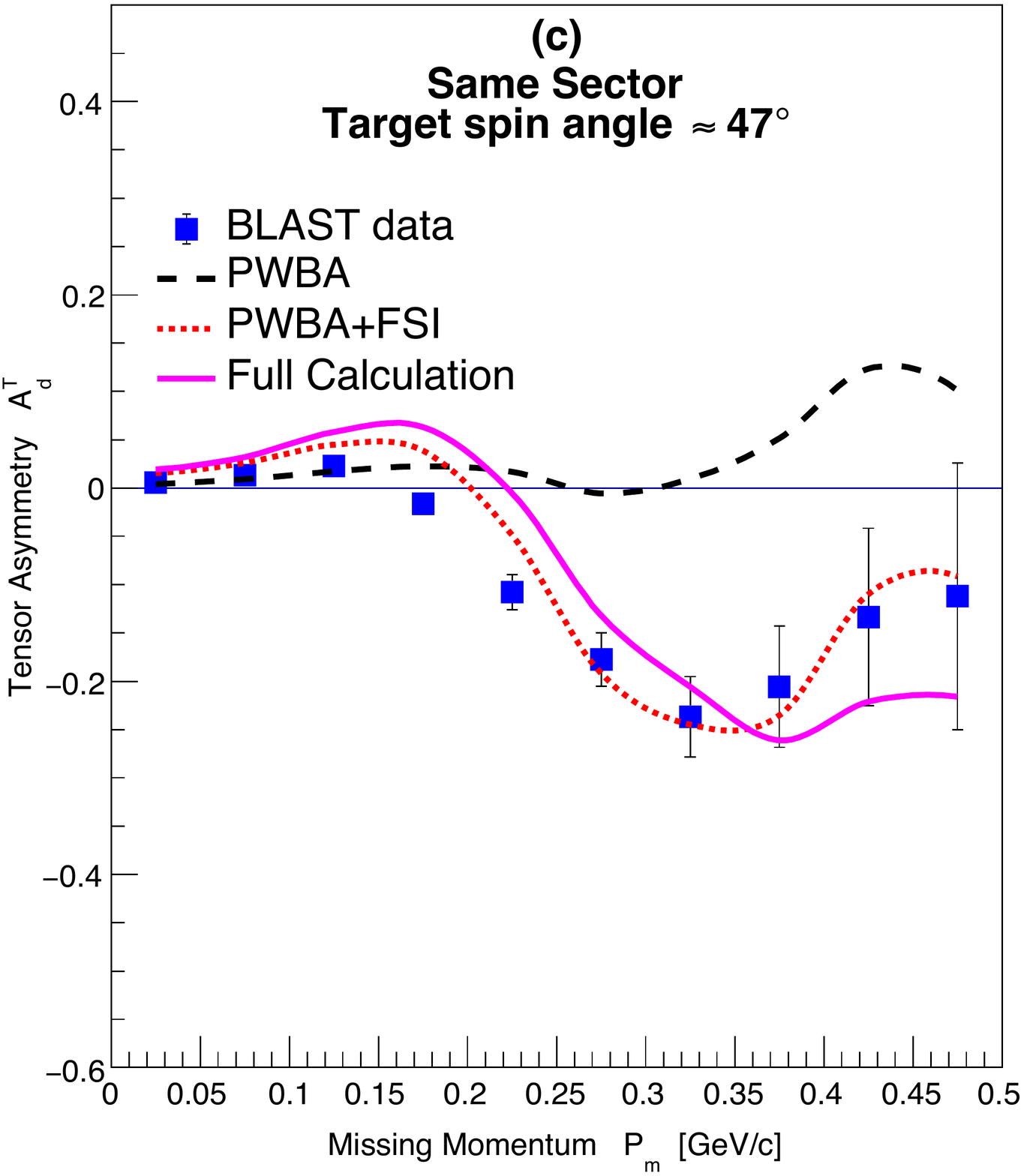}
\hfil
\includegraphics[width=0.246\textwidth, viewport=0 45 562 690, clip]{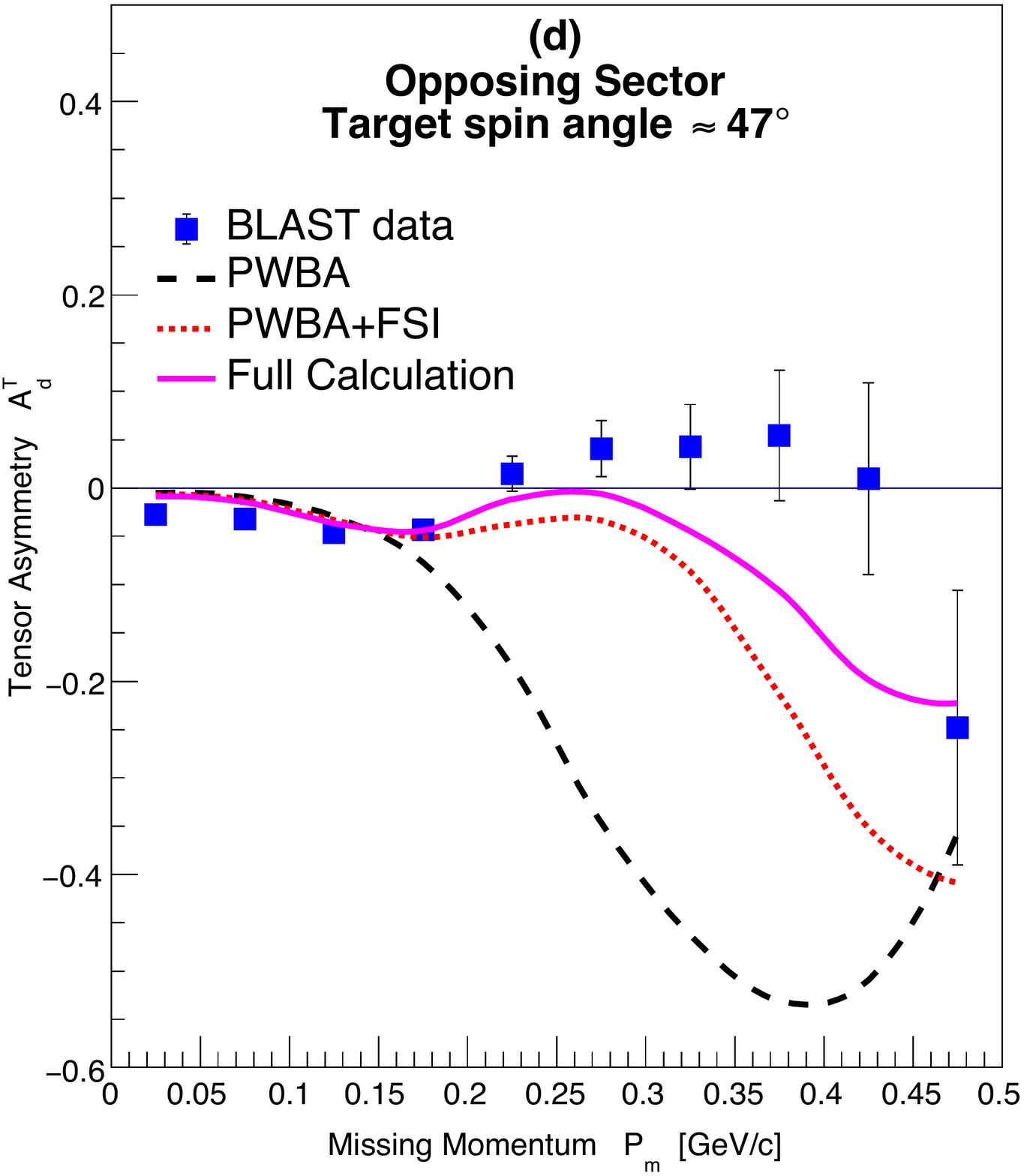}
\caption{Tensor asymmetries $A^T_d$ for $0.1< Q^2< 0.5$~(GeV/c)$^2$
  vs. $p_m$.  Panels (a) and (c) refer to {\it same sector} kinematics
  for target spin angles $\approx32^\circ$ and $\approx47^\circ$.
  Panels (b) and (d) refer to {\it opposing sector} kinematics for the
  same target spin angles.}
\label{fig:tensor}
\end{figure*} 
Figure~\ref{fig:vector} shows that the experimental asymmetries
$A^V_{ed}$ are in good agreement with the full theoretical
calculations over a wide range of $Q^2$ and $p_m$. The only previous
measurement of $A^V_{ed}$ was carried out in perpendicular kinematics
at NIKHEF~\cite{NIKHEF:2002} up to $p_m$ of about 300~MeV/c, although
with limited statical precision after 200~MeV/c. The BLAST data in the
region of $p_m$ around 200~MeV/c, where the $S$- and $D$-states
strongly interfere, are very well described by the full theoretical
calculation in contrast to the claim in~\cite{NIKHEF:2002} where the
data suggested an underestimation by the theory.  The $A^V_{ed}$
asymmetries directly relate to the deuteron momentum distribution for
the $M_d = \pm 1$ spin states. It has been pointed
out~\cite{Forest1996} that the $M_d = 1$ momentum distribution has a
zero around 300~MeV/c, which in a simple picture can be related to the
dimensions of the toroidal shape of the density distribution. The
Fourier transform of the deuteron density calculated in the model
of Ref.~\cite{Forest1996} yields a zero at 320~MeV/c for the $M_d = 1$
momentum distribution~\cite{DeGrush}, which is where the $A^V_{ed}$
asymmetries in Fig.~\ref{fig:vector} have their zero-crossings. This
zero crossing was also predicted by Jeschonnek and
Donnelly~\cite{Donnelly1998} using an improved treatment of the
non-relativistic reduction of the electromagnetic current operator.

Figure~\ref{fig:tensor} shows the tensor analyzing powers $A^T_d$ as a
function of $p_m$ for the same kinematics and target spin orientations
as that of Fig.~\ref{fig:vector}, and compared also with the same
theoretical model folded with the detector acceptances and
efficiencies. Just as for $A^V_{ed}$ the only previous measurement of
$A^T_{d}$ was carried out in parallel kinematics at
NIKHEF~\cite{NIKHEF:1999} up to $p_m$ of only 150~MeV/c with limited
statistics. The BLAST $A^T_d$ data extend up to $p_m$ = 500~MeV/c and
for the first time into the region where the $D$-state dominates over
the $S$-state. As expected, where the $S$-state dominates, the $A^T_d$
are small and well described by the theoretical calculations,
including the simple PWBA.  Beyond about $p_m$ = 150~MeV/c $A^T_d$
grows, indicating the effect of the tensor polarization. The PWBA
calculations show that the sign is different for the $A^T_d$ in {\it
  same sector} and {\it opposing sector} kinematics.

As shown in Fig.~\ref{fig:tensor}, in contrast to the vector
asymmetries $A^V_{ed}$, the tensor asymmetries $A^T_d$ are significantly
modified by the effects of the FSI for $p_m \geq 150$~MeV/c. In {\it
  same sector} kinematics, the effects of FSI bring the $A^T_{d}$
calculations into reasonable agreement with the present data. In {\it
  opposing sector} kinematics, the effects of the FSI are also sizable
but not sufficient to agree with the data; the effects of MEC and IC
contribute equally after FSI to produce the full calculations of
Fig.~\ref{fig:tensor}. The kinematic reach of the BLAST data is such
that the proton-neutron interaction is sampled via the FSI over a
large spatial range: from short distances, where the nucleons are
expected to overlap, to long distances where the interaction is
dominated by one-pion-exchange. The $A^T_d$ data at
$p_m\geq 250$~MeV/c are particularly sensitive to the tensor part of
the interaction at short distances, where it has significant
model dependence~\cite{Forest1996}. It is to be noted that the
theoretical model used here works well, given that it is mainly based
on nucleon degrees of freedom.

We have presented data for the vector $A^V_{ed}$ and tensor $A^T_d$
spin asymmetries from the deuteron for $0.1<Q^2<0.5$ ~(GeV/c)$^2$.
The asymmetries were mapped out for quasielastic kinematics
$(\vec{e}, e^{\prime}p)$ over a range of $p_m$ up to $\sim500$~MeV/c.
The data were taken using an internal deuterium gas target polarized
in both vector and tensor spin states that minimized systematic
errors. This was done simultaneously with precise measurements of the
elastic~\cite{Zhang:2011} and the $(e, e^{\prime}n)$~\cite{Geis:2008}
channels that also permitted measurements of $P_{zz}$ and $P_z$.  The
new data are in good agreement with theoretical calculations and
provide a strong constraint on our understanding of deuteron structure
and the tensor force between a neutron and a proton.  The $D$-state
contribution is clearly evident in both asymmetries as $p_m$ increases
and highlights the importance of measurements at large $p_m$.  The
tensor asymmetries with same and opposing sector kinematics probes the
proton-neutron interaction over a large spatial range.  These results
and approach are important for future theoretical calculations and
experiments that study the deuteron and details of the proton-neutron
interaction.

\begin{acknowledgments}
We thank the staff at the MIT-Bates Linear Accelerator Center for the
high quality electron beam and their technical support. We thank H.
Arenh\"ovel for many enlightening discussions. This work has been
supported in part by the US Department of Energy Office of Nuclear
Physics and by the National Science Foundation.
\end{acknowledgments}

\end{document}